\documentclass[journal=jpccck,manuscript=article]{achemso}

\usepackage[T1]{fontenc}       % Use modern font encodings

\usepackage{bm}
\usepackage{bbm}
\usepackage{mathrsfs}
\usepackage{color}
\usepackage{graphicx}
\usepackage{dcolumn}
\usepackage{amssymb}
\usepackage{amsmath}

\title{Evaluating alignment of elongated nanoparticles in cylindrical geometries through small angle X-ray scattering experiments}% Force line breaks with \\
\author{Tomas Ros\'{e}n}
\email{rosen@mech.kth.se}
\affiliation{KTH Mechanics, Qsquars backe 18,Royal Institute of Technology, SE-100 44 Stockholm, Sweden}
\alsoaffiliation{Wallenberg Wood Science Center, Royal Institute of Technology, SE-100 44 Stockholm, Sweden}

\author{Christophe Brouzet}
\affiliation{KTH Mechanics, Qsquars backe 18,Royal Institute of Technology, SE-100 44 Stockholm, Sweden}

\author{Stephan V. Roth}
\affiliation{DESY, Notkestrasse 85, Hamburg, Germany}
\alsoaffiliation{Department of Fibre and Polymer Technology, Teknikringen 56-58, Royal Institute of Technology, SE-100 44 Stockholm, Sweden}

\author{Fredrik Lundell}
\affiliation{KTH Mechanics, Qsquars backe 18,Royal Institute of Technology, SE-100 44 Stockholm, Sweden}
\alsoaffiliation{Wallenberg Wood Science Center, Royal Institute of Technology, SE-100 44 Stockholm, Sweden}

\author{L. Daniel S\"oderberg}
\affiliation{KTH Mechanics, Qsquars backe 18,Royal Institute of Technology, SE-100 44 Stockholm, Sweden}
\alsoaffiliation{Wallenberg Wood Science Center, Royal Institute of Technology, SE-100 44 Stockholm, Sweden}

\date{\today}% It is always \today, today,
\begin{document}

\begin{abstract}
The increased availability and brilliance of new X-ray facilities have in the recent years opened up the possibility to characterize the motion of dispersed nanoparticles in various microfluidic applications. One of these applications is the process of making strong continuous filaments through hydrodynamic alignment and assembly of cellulose nanofibrils (CNF) demonstrated by H\aa kansson~{et al.~}[Nature communications \textbf{5}, 2014]. In this process it is vital to study the alignment of the nanofibrils in the flow, as this in turn affects the final material properties of the dried filament. Small angle X-ray scattering (SAXS) is a well-suited characterization technique for this, which typically provides the alignment in a projected plane perpendicular to the beam direction. In this work, we demonstrate a simple method to reconstruct the full three-dimensional (3D) orientation distribution function (ODF) from a SAXS-experiment through the assumption that the azimuthal angle of the nanofibril around the flow direction is distributed uniformly; an assumption that is approximately valid in the flow-focusing process. For demonstrational purposes, the experimental results from H\aa kansson~{et al.~}(2014) have been revised, resulting in a small correction to the presented order parameters. The results are then directly compared with simple numerical models to describe the increased alignment of CNF both in the flowing system and during the drying process. The proposed reconstruction method will allow for further improvements of theoretical or numerical simulations and consequently open up new possibilities for optimizing assembly processes, which include flow-alignment of elongated nanoparticles.
\end{abstract}

\maketitle

\begin{figure}
\includegraphics[width=0.99\textwidth]{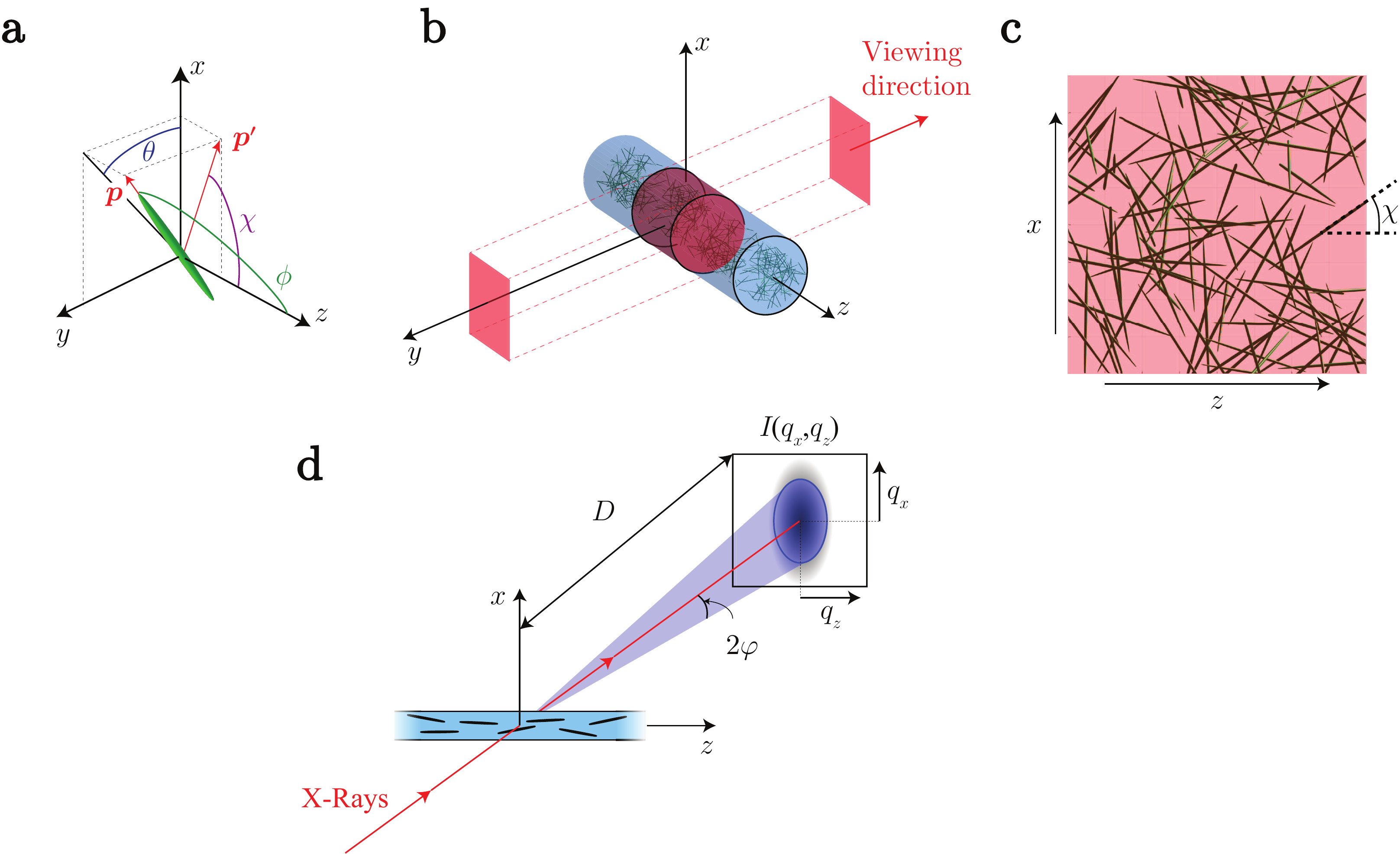}% Here is how to import EPS art
\caption{(a) Definition of the orientation of an elongated particle with unit vector $\boldsymbol{p}$ along the major axis; the polar angle of the particle to the $z$-direction is denoted with $\phi$ and the corresponding azimuthal angle in the $xy$-plane is $\theta$; the projection of the unit vector $\boldsymbol{p}$ on the viewing ($xz$) plane is denoted $\boldsymbol{p}'$ and has the angle $\chi$ with respect to the $z$-axis; (b) illustration of a flow in a typical experiment; the flow is in $z$-direction, the viewing direction is defined in negative $y$-direction; (c) a typical projection of the elongated particles in the viewing plane; (d) schematic illustration of a typical SAXS experiment.}
  \label{fig:Paper8FlowProblem}
\end{figure}

\section{\label{sec:Paper8Intro} Introduction}
Studying the alignment of anisotropic nanoparticles or polymers in flows is relevant in many applications, for example in fibre spinning \cite{yang1994fiber} or creating nanostructured films \cite{long2012recent,blell2016generating}. In many of these material processes, the particle shapes and orientations will have a significant effect on macroscopic properties. For example, the macroscopic electrical response of films made from carbon nanotubes is greatly affected by the alignment of the nanoparticles inside the material \cite{behnam2007,allen2012,allen2013,shekhar2011,simoneau2013}. Other applications relate to the influence of the orientation on macroscopic optical properties of the material such as its refractive index \cite{blell2016generating}. 

The orientation distribution of the anisotropic nanoparticles inside the material is often dependent on their dynamics in a flowing dispersed state in the process. Therefore it is of great importance to understand how the angular dynamics and alignment of nanoparticles in flows can be studied, modeled and controlled during the processing of materials. It is known that nanoparticles typically align from shear and extensional flow \cite{trebbin2013anisotropic,lutz2016scanning}. However, since shear also causes rotation of particles, it can be preferable to align particles only with extensional flow \cite{Hakansson_NatComm}.

The description of the particle orientation in the flow is usually through spherical coordinates with polar angle $\phi$ between the particle major axis and the flow ($z$) direction, and the azimuthal angle $\theta$ in the plane perpendicular to the flow (see figure~\ref{fig:Paper8FlowProblem}a-b). Here, the viewing direction is the negative $y$-direction, such that the viewing plane is equal to the $xz$-plane. In that plane we define the particle orientation with the projected angle $\chi$. As seen in the figure, these angles are related through\\
\begin{equation}
\tan\chi=\tan\phi\cos\theta.
\label{eq:projectedAngle}
\end{equation}
~\\
The nanometer-sized particles of interest are smaller than the visible wavelength of light, and the understanding of their shape and properties must rely on using other characterization techniques than standard microscopy. In solid material, there is the possibility to use highly magnified scanning/transmission electron microscopy (SEM/TEM) images  or atomic force microscopy (AFM) \cite{hanley1992}. However, to present any statistically relevant orientation distribution function (ODF) describing the collective orientation of particles would require a substantial amount of images being analyzed. Furthermore, we will only obtain the projected angle $\chi$ on the viewing plane from these images (see figure~\ref{fig:Paper8FlowProblem}c) unless the nanoparticles are mono-dispersed and the out-of-plane orientation can be found through the length of the projected particle. Due to the nature of the techniques, neither SEM nor AFM are suitable for studying flowing systems of dispersed nanoparticles.

Following the increased availability and performance of synchrotron light sources, small angle X-ray scattering (SAXS) has proven to be a vital tool for nanoparticle characterization in flowing dispersions (see figure~\ref{fig:Paper8FlowProblem}d) \cite{trebbin2013anisotropic,lutz2016scanning,ghazal2016}. With this technique, the light scattered by the particles is recorded on a detector and this information will describe the shape and size of the particles in an isotropic system. If the particles are non-spherical and show some preferential alignment, the technique will also provide a statistically relevant ODF \cite{stribeck}. However, when using SAXS in a transmission geometry, it is important to take into consideration that the experiment just provides the distribution of the projected angle $\chi$. Furthermore, the ODF of the polar angle $\phi$ is not easy to obtain from the projected distribution of $\chi$. This particular issue will result in a non-trivial comparison between experiments and simulations when it comes to describing the angular dynamics of nanoparticles in flows. To obtain the three-dimensional ODF with SAXS, the typical approach is by rotating the sample and thus obtaining the projected ODF from different incidence angles from which the three-dimensional orientation can be reconstructed during post-processing \cite{wagermaier2006spiral,wagermaier2007scanning,schaff2015six,skjonsfjell2016}.

\begin{figure}
\includegraphics[width=0.99\textwidth]{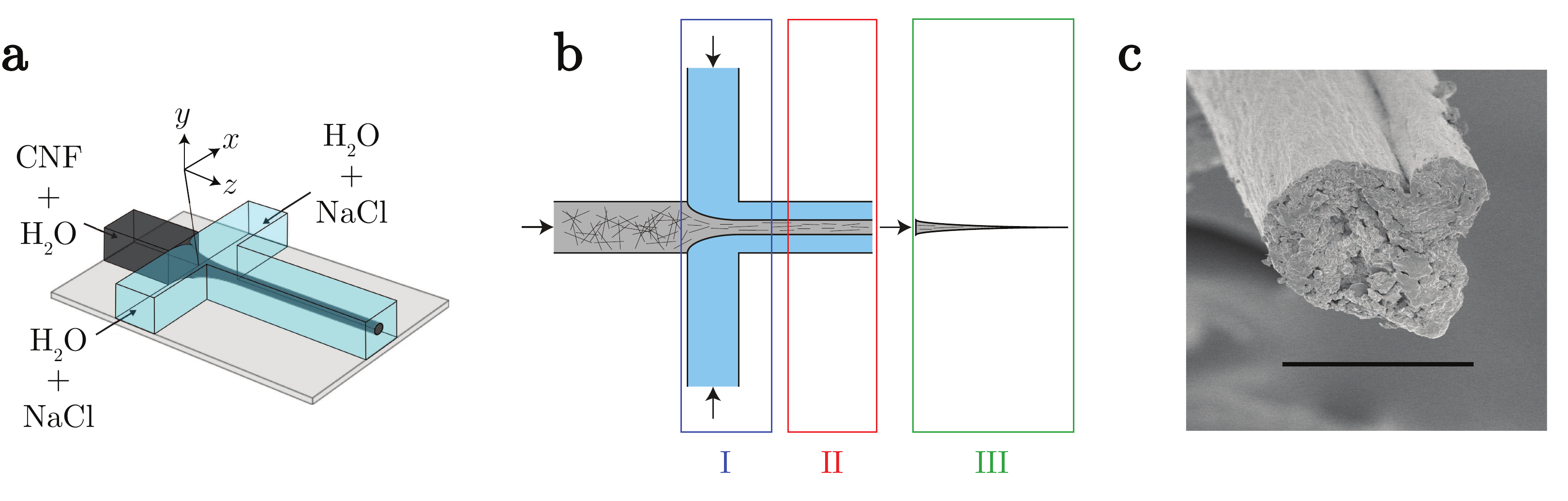}% Here is how to import EPS art
\caption{Demonstration of flow-focusing device and the process to make strong cellulose filaments from CNF; (a) the flow-focusing device with a core flow of a CNF suspension and sheath flows containing salt and water; (b) the filament production consists of three steps: (I) hydrodynamic alignment of CNF in the $z$-direction, (II) transition of the CNF from a dispersion to an arrested (gel) state, (III) drying of the gel-thread; (c) SEM image of the final dried cellulose filament with the black scale bar indicating $20$~$\mu$m; figures adapted from \citet{Hakansson_NatComm}.}
  \label{fig:Paper8FlowFocusing}
\end{figure}

\subsection{Making new materials from cellulose}

Structural materials in nature are found to have properties that can surpass the performance of their individual components\cite{wegst2015bioinspired}, and thus offer great potential to meet the demands of a sustainable society. Using biobased nanoparticles as building blocks, there have recently been several examples of using flow alignment to control the assembly of new types of materials and composites\cite{Hakansson_NatComm,hamedi2014highly,mittal2017ultrastrong,kamada2017flow}. Using a dispersion of cellulose nanofibrils (CNF) \citet{Hakansson_NatComm} demonstrated a process of making strong and stiff cellulose filaments that could potentially be used in new biobased composites or high performance textiles. This process is schematically illustrated in figure~\ref{fig:Paper8FlowFocusing} and will be described shortly. In a flow-focusing device, a dispersion of CNF is focused by a sheath flow of a NaCl solution. During the acceleration of the core flow, the fibrils are hydrodynamically aligned in the flow direction almost solely due to extensional flow. As the ions in the sheath flow are diffusing into the core, the CNF is forming a gel where the internal structure of fibrils are locked in an aligned state. After a subsequent drying of the gel thread, a continuous filament is produced with strength and stiffness that could potentially be comparable to glass fibers. The mechanical properties of the resulting filament were seen to be dependent on the degree of alignment of CNF along the filament direction. Increasing the alignment of CNF in the flow-focusing process could thus potentially lead to an even stronger material. The alignment of the CNF in the flow was studied by \citet{Hakansson_NatComm} using SAXS by quantifying the anisotropy in the scattering pattern.

In a later work, \citet{Hakansson_SAXSALIGN} demonstrated how the evolution of the ODF of particles in the flow-focusing device can be simulated. The simulated alignment based on the polar angle $\phi$ was then compared with the alignment based on the projected angle $\chi$ in the SAXS patterns in a simplified manner.

Another crucial aspect of the process is the drying of the gel thread. When water is removed from the gel, the spatial distribution of fibers is assumed to shrink radially. This radial shrinking is also believed to cause further alignment as the projected radial component of the particle symmetry axis also will decrease with the same rate. When performing scattering experiments of a dried cellulose filament, we are thus likely to measure a higher degree of alignment than what is measured in the channel. \citet{Hakansson_NatComm} characterized the alignment of the nanofibrils in the dried filament with wide angle X-ray scattering (WAXS), relying on the same principle as SAXS. This means that the mean fibril angle obtained in this study also represents the mean projected angle $\langle\chi\rangle$ and not the mean polar angle $\langle\phi\rangle$ to the fiber direction. 

\subsection{Objective of the present work}

In this work, it will be shown how the alignment based on the mean polar angle in the flow is related to the projected alignment in the viewing plane. The only assumption that is made when deriving this relationship is that the flow has cylindrical symmetry and therefore that the distribution of the azimuthal angle $\theta$ in the plane perpendicular to the flow direction is uniform. This assumption was demonstrated to be also valid in the flow-focusing process of CNF by \citet{Hakansson_NatComm}. It will also be discussed what will happen to the ODF during the drying process. It is found here that the measured (projected) alignment could be significantly lower than the alignment based on the polar angle and the discrepancy depends heavily on the type of flow or other alignment mechanism that the fibrils are subject to. However, when applying the reconstruction algorithm to obtain the ODF based on $\phi$ from the SAXS/WAXS data by \citet{Hakansson_NatComm}, it is found that the particular error for their results still is small.

The present study will have the following outline. Firstly, some background information will be provided to the reader regarding some basic techniques to study the angular dynamics of dispersed nanofibrils both numerically and experimentally using SAXS. In the following sections, the new reconstruction method will be presented and validated. Finally, in the last sections, the reconstruction method will be applied to the data by \citet{Hakansson_NatComm} along with a comparison with simple numerical models.

\section{Theory}
\subsection{Dynamics of small particles in flows}

The probability of a fibril to have an orientation in the interval $\phi\in[\phi_1,\phi_2]$ and $\theta\in[\theta_1,\theta_2]$ is given by the following expression using the ODFs $\Psi_\phi$ and $\Psi_\theta$\\
\begin{equation}
\text{Pr}[\phi_1\leq\phi\leq\phi_2,\theta_1\leq\theta\leq\theta_2]=\int_{\theta_1}^{\theta_2}\Psi_\theta d\theta\int_{\phi_1}^{\phi_2}\Psi_\phi|\sin\phi|d\phi,
\label{eq:Probability}
\end{equation}
~\\
where the two functions are normalized according to
\begin{equation}
2\pi\int_{-\pi/2}^{\pi/2}\Psi_\phi|\sin\phi|d\phi=1\\
\label{eq:Probability2}
\end{equation}
\begin{equation}
\frac{1}{2\pi}\int_{0}^{2\pi}\Psi_\theta d\theta=1.\\
\label{eq:Probability3}
\end{equation}

~\\

Throughout this work, it is going to be assumed that the azimuthal angle $\theta$ is distributed uniformly, i.e.~$\Psi_\theta=1$.

In the flow-focusing process illustrated in figure~\ref{fig:Paper8FlowFocusing}, it is reasonable to assume that the core flow of CNF is only in the $z$-direction and the velocity $\boldsymbol{u}=(u,v,w)$ does not vary in the radial direction, i.e.~$ u=v=0$\cite{Hakansson_SAXSALIGN}. This means that $\Psi_\theta$ always will remain constant and $\Psi_\phi$ will only be dependent on the centerline velocity $w_c=w_c(z)$, the extension rate $\dot{\varepsilon}=\dot{\varepsilon}(z)=dw_c/dz$ and the rotary diffusion coefficient $D_r$. Assuming that the dispersion is dilute and the fibrils do not interact with each other and further assuming that the fibrils can be described as stiff spheroids with aspect ratio $r_p$ (major axis/minor axis), the steady state of $\Psi_\phi=\Psi_\phi(z)$ is given by the Smoluchowski equation\cite{doi1986theory,Hakansson_SAXSALIGN}\\
\begin{equation}
w_c\frac{\partial \Psi_\phi}{\partial z}=\frac{1}{\sin{\phi}}\frac{\partial}{\partial \phi}\left(\underbrace{D_r\sin\phi\frac{\partial \Psi_\phi}{\partial \phi}}_\text{rotary diffusion}-\underbrace{\sin\phi\frac{\partial\phi}{\partial t}\Psi_\phi}_\text{hydrodynamic forcing}\right),
\label{eq:CylindricalSmoluchowski}
\end{equation}
~\\
where the angular velocity is determined by \citet{Jeffery} as:\\
\begin{equation}
\frac{\partial\phi}{\partial t} =-\dot{\varepsilon}\Lambda\frac{3}{2}\cos\phi\sin\phi
\end{equation}
with
\begin{equation}
\Lambda=\frac{r_p^2-1}{r_p^2+1}.
\end{equation}
~\\
The assumptions made to justify this particular model are of course very questionable for the actual flow of CNF under process-relevant conditions, but will be used in this work only for the sake of discussion. The details about the derivation of this model along with improvements to account for interacting fibrils are given by \citet{doi1986theory} and \citet{Hakansson_SAXSALIGN}.

The Smoluchowski equation above is only dependent on one dimensionless parameter, called the P\'{e}clet number $Pe$ and is defined as:\\
\begin{equation}
Pe=\frac{\dot{\varepsilon}\Lambda}{D_r},
\end{equation}
~\\
which relates the effect of hydrodynamic forcing with the effect of Brownian rotary diffusion. For example, if $Pe=\infty$, the rotary diffusion term can be neglected in eq.~(\ref{eq:CylindricalSmoluchowski}). Analogously, if $Pe=0$, the hydrodynamic forcing term can be neglected. For a constant value of $Pe$ (constant extension rate and rotary diffusion), there will be an equilibrium distribution given by the solution of $d\Psi_\phi/dz=0$. The equilibrium distribution can be analytically found to be (see supplementary information for full derivation):\\
\begin{equation}
\Psi_{\phi}^\text{eq}=\frac{\exp(-\frac{\sin^2 \phi}{2\sigma^2})}{\int_0^{2\pi}d\theta\int_{-\pi/2}^{\pi/2} \exp(-\frac{\sin^2 \phi}{2\sigma^2})|\sin \phi| \textrm{d}\phi,}
\label{eq:equlibriumDistributions}
\end{equation}
with
\begin{equation}
\sigma^2=\frac{2}{3Pe}.
\end{equation}

\subsection{\label{sec:Paper8Alignment} The order parameter}
To describe the collective alignment of many elongated particles in the flow, the mean of the second Legendre polynomial $S_\phi=\langle P_2(\cos\phi)\rangle$ is commonly used and sometimes called \emph{Hermans order parameter}. This quantity is expressed as \cite{vanGurp}:\\
\begin{equation}
S_\phi=\langle P_2(\cos\phi)\rangle=\left\langle\frac{3}{2}\cos^2\phi-\frac{1}{2}\right\rangle,
\end{equation}
~\\
where the brackets denote an ensemble average over all particles. Consequently, $S_\phi=0$ if all particles are uniformly distributed and $S_\phi=1$ if the particles are perfectly aligned in the flow direction. Using the ODF $\Psi_\phi$, the order parameter $S_\phi$ can be obtained through\\
\begin{equation}
S_\phi=\int_0^{2\pi}d\theta\int_{-\pi/2}^{\pi/2}\Psi_\phi\left(\frac{3}{2}\cos^2\phi-\frac{1}{2}\right)|\sin\phi|d\phi.
\end{equation} 
~\\
Additionally, it should be mentioned that the value of $S_\phi$ might not always be a good measure of the true ODF $\Psi_\phi$ since it corresponds to an integrated quantity. This means of course that different ODFs can correspond to the same order parameter. For example, an order parameter of $S_\phi=0$ could both mean that the system is isotropic, but could also mean that all fibers are perfectly oriented at an angle of $\phi=\arccos\sqrt{1/3}\approx 54.7^\circ$.

\begin{figure}[t]
\includegraphics[width=0.99\textwidth]{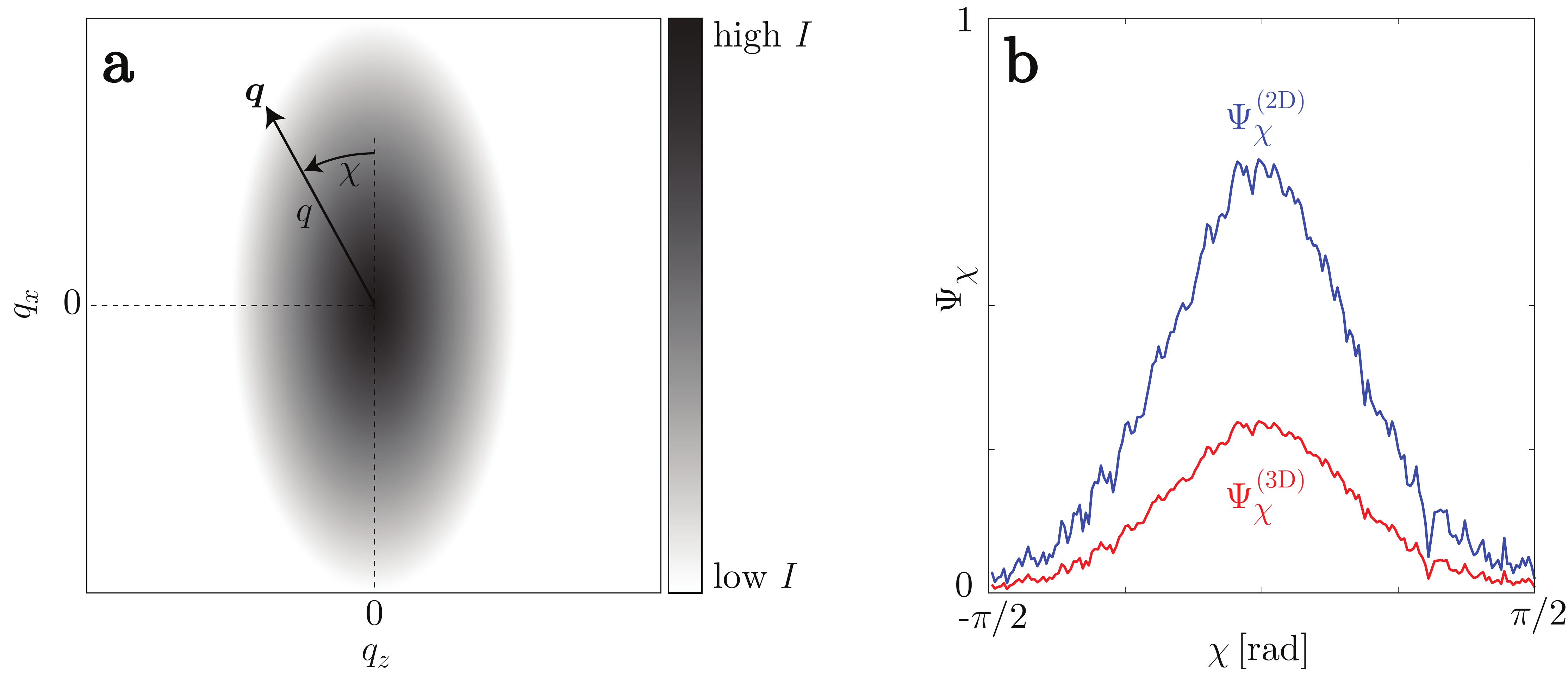}% Here is how to import EPS art
\caption{(a) A {\bf \underline{schematic}} illustration of the scattering intensity on the detector in a SAXS experiment; (b) the averaged ODF $\Psi_\chi$ obtained from the SAXS experiments by \citet{Hakansson_NatComm} in the flow-focusing device at $z=2.5h$; the blue curve shows the ODF $\Psi_\chi$  normalized using eq.~(\ref{eq:2DNorm}) and is denoted $\Psi_\chi^\text{(2D)}$ while the red curve shows the ODF normalized with eq.~(\ref{eq:3DNorm}) and is denoted $\Psi_\chi^\text{(3D)}$.}
  \label{fig:Paper8SAXSPrinciple}
\end{figure}

\subsection{SAXS experiments}
The typical setup for a transmission SAXS experiment used by \citet{Hakansson_NatComm} to study the alignment of nanofibrils in a flow was illustrated previously in figure~\ref{fig:Paper8FlowProblem}d. The X-ray beam with wavelength $\lambda$ is illuminating the flowing dispersion in the negative $y$-direction, perpendicular to the flow ($z$) direction. Some of the incoming photons are scattered by the particles and the scattered light is a representation in the reciprocal space of the illuminated fibrils. The scattered light intensity $I$ is collected on a detector placed at a distance $D$ from the sample. A schematic illustration of the X-ray scattering intensity on the detector is given in figure~\ref{fig:Paper8SAXSPrinciple}a. The anisotropy in the SAXS pattern, with higher scattering intensity in the $x$-direction, reflects the fact that the sample is preferentially oriented in the $z$-direction. After background subtraction, the normalized intensity along a constant $q=(4\pi/\lambda)\sin\varphi$ provides an estimate of the ODF $\Psi_\chi$ of the projected fibril angle $\chi$.

Typically, there are two ways often used to normalize the function $\Psi_\chi$. Either the angle~$\chi$ is treated as an azimuthal angle in a 2D plane and is thus normalized according to 
\begin{equation}
\int_{-\pi/2}^{\pi/2} \Psi^\text{(2D)}_\chi \textrm{d}\chi =1,
\label{eq:2DNorm}
\end{equation}
~\\
or the angle $\chi$ is treated as a representation of the polar angle in 3D space and thus normalized according to
\begin{equation}
\int_0^{2\pi}d\theta\int_{-\pi/2}^{\pi/2} \Psi^\text{(3D)}_\chi |\sin \chi| \textrm{d}\chi =1.
\label{eq:3DNorm}
\end{equation}
~\\
The two distributions $\Psi^\text{(3D)}_\chi $ and $\Psi^\text{(2D)}_\chi$ are only differing with a constant scaling factor $\alpha$ such that $\Psi^\text{(3D)}_\chi=\alpha \Psi^\text{(2D)}_\chi$ with
\begin{equation}
\alpha=\left(\int_0^{2\pi}d\theta\int_{-\pi/2}^{\pi/2} \Psi^\text{(2D)}_\chi |\sin \chi| \textrm{d}\chi \right)^{-1}.
\end{equation}
~\\
An illustration showing the difference in scaling of $\Psi_\chi$ is illustrated in figure~\ref{fig:Paper8SAXSPrinciple}b.

In the work by \citet{Hakansson_NatComm}, the order parameter obtained from the SAXS experiments is found by using the 3D normalization of $\Psi_\chi$ according to\\
\begin{equation}
S_\chi=\int_0^{2\pi}d\theta\int_{-\pi/2}^{\pi/2}\Psi^\text{(3D)}_\chi\left(\frac{3}{2}\cos^2\chi-\frac{1}{2}\right)|\sin\chi| \textrm{d}\chi.
\label{eq:Paper8schi}
\end{equation}
~\\
Given this definition, $S_\chi$ is also equal to zero when the particles have an isotropic orientation distribution and $S_\chi=1$ if the particles are perfectly aligned in $z$-direction. However, for an arbitrary orientation distribution the two order parameters $S_\chi$ (based on the projected angle~$\chi$) and $S_\phi$ (based on the polar angle~$\phi$) are \underline{not} the same.

\section{Method}
\subsection{\label{sec:3Dto2D} Reconstructing $\Psi_\phi$ from $\Psi_\chi$}
The SAXS experiments give us access to the projected ODF $\Psi_\chi$ of the ODF $\Psi_\phi$. Depending on normalization of $\Psi_\chi$ according to eqs.~(\ref{eq:2DNorm})~and~(\ref{eq:3DNorm}), the ODFs $\Psi_\chi^\text{(2D)}$ and $\Psi_\chi^\text{(3D)}$ can be obtained. To be able to reconstruct $\Psi_\phi$ from $\Psi_\chi$, we can use the different symmetries of the system. We assume here that the ODF is axi-symmetric around the flow axis $z$, meaning that the ODF depends only on $\phi$ and not on $\theta$. Following~\citet{vanGurp}, the ODF $\Psi_\phi$ can be expressed as a series expansion of  Legendre polynomials:\\
\begin{equation}
\Psi_\phi=\sum_{j=0,2,4,...}^{2(N_\text{LP}-1)}\frac{2j+1}{4\pi} \langle P_j \rangle_\phi P_j(\cos \phi),
\label{eq:legendre_decomposition}
\end{equation}
~\\
where $P_j$ is the Legendre polynomial of order $j$ and $\langle P_j \rangle_\phi=\langle P_j(\cos\phi) \rangle$ is the $j$-th order parameter of the ODF. The exact result is obtained as $N_\text{LP}\rightarrow\infty$. Note that $S_\phi$ corresponds to the second order parameter $\langle P_2 \rangle_\phi$. Here, we use only the even Legendre polynomials because the ODF follows the symmetry condition $\Psi_\phi(\phi)=\Psi_\phi(\pi-\phi)$.

In the same way, the projected ODF using the 3D normalization can also be decomposed in Legendre polynomials:\\
\begin{equation}
\Psi^\text{(3D)}_\chi=\sum_{i=0,2,4,...}^{2(N_\text{LP}-1)}\frac{2i+1}{4\pi} \langle P_i \rangle_\chi P_i(\cos \chi),
\label{eq:legendre_decomposition_chi}
\end{equation}
~\\
with $\langle P_i \rangle_\chi=\langle P_i(\cos\chi) \rangle$. 

The order parameters $\langle P_j \rangle_\phi$ and $\langle P_i \rangle_\chi$ are related through (see derivation in supplementary information)\\
\begin{equation}
\left(
\begin{matrix}
\langle P_0 \rangle_\chi\\
\langle P_2 \rangle_\chi\\
\vdots\\
\langle P_N \rangle_\chi\\
\end{matrix}
\right)
=
\frac{\alpha}{2}
\left(
\begin{matrix}
C_{0,0} & C_{0,2}&\hdots&C_{0,N}\\
C_{2,0} & C_{2,2}&\hdots&C_{2,N}\\
\vdots & \vdots & \ddots & \vdots\\
C_{N,0} & C_{N,2}&\hdots& C_{N,N}
\end{matrix}
\right)
\left(
\begin{matrix}
\langle P_0 \rangle_\phi\\
5\langle P_2 \rangle_\phi\\
\vdots\\
(2N+1)\langle P_{N} \rangle_\phi\\
\end{matrix}
\right)
\label{eq:link_between_orderparameters}
\end{equation}
~\\
with $N=2(N_\text{LP}-1)$ and the $N_\text{LP}\times N_\text{LP}$ matrix $C_{i,j}$ given by\\
\begin{equation}
C_{i,j}= \int_{-\pi/2}^{\pi/2}P_i(\cos \chi) |\sin \chi| \left(\int_{0}^{1} \frac{4 p'}{\sqrt{1-{p'}^2}}  P_j(p' \cos \chi) \textrm{d}p'\right) \textrm{d}\chi,\\
\label{eq:TheMatrix_C}
\end{equation}
~\\
for $i,j=0,2,4~...~2(N_\text{LP}-1)$. Note here that the matrix $C_{i,j}$ does not depend on the actual ODF, and can be pre-computed for a given $N_\text{LP}$. 

Now, the ODF $\Psi_\phi$ based on the polar angle can be reconstructed from the projected ODF $\Psi_\chi$ using the following steps:
\begin{enumerate}

\item Normalize $\Psi_\chi$ in two ways according to eqs.~(\ref{eq:2DNorm})~and~(\ref{eq:3DNorm}) to obtain $\Psi^\text{(2D)}_\chi$, $\Psi^\text{(3D)}_\chi$~and~$\alpha$.

\item Find the order parameters $\langle P_i \rangle_\chi$ using
\begin{equation}
\langle P_i \rangle_\chi =  \int_0^{2\pi}d\theta\int_{-\pi/2}^{\pi/2} \Psi^\text{(3D)}_\chi P_i(\cos \chi) |\sin \chi| \textrm{d}\chi
\end{equation}

\item Compute the matrix $C_{i,j}$ for the~$N_\text{LP}$ first even Legendre polynomials using eq.~(\ref{eq:TheMatrix_C}). The exact analytical values of the matrix $C_{i,j}$ for $N_\text{LP}=15$ is given as a separate file in the online supplementary material.

\item Inverse eq.~(\ref{eq:link_between_orderparameters}) to obtain the order parameters $\langle P_j \rangle_\phi$.

\item Reconstruct $\Psi_\phi$ using eq.~(\ref{eq:legendre_decomposition}).

\end{enumerate}

\begin{figure}[t]
\includegraphics[width=0.99\textwidth]{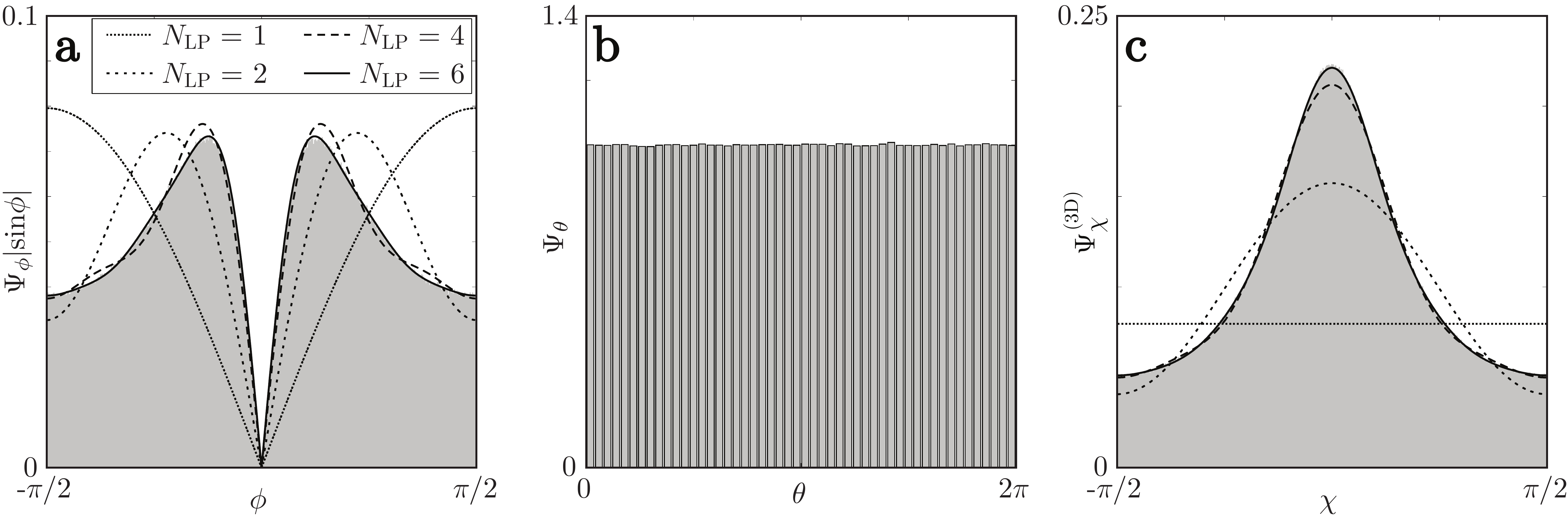}% Here is how to import EPS art
\caption{Histogram for $N=10^7$ particles illustrated with the gray area; (a) the ODF $\Psi_\phi$ obtained from the Smoluchowski simulations with uni-axial extension and no rotary diffusion at $S_\phi=0.30$; (b) the angle $\theta$ is sampled from a uniform distribution; (c) the resulting ODF $\Psi_\chi^\text{(3D)}$ with a projected order parameter of $S_\chi=0.19$; the curves in (a) and (c) correspond to the reconstruction of $\Psi_\phi$ for given $N_\text{LP}$ of Legendre polynomials used to fit $\Psi_\chi^\text{(3D)}$ in (c).}
  \label{fig:Paper8angleHistograms}
\end{figure}

\section{\label{sec:Paper8Validation} Validation of the reconstruction method}
To demonstrate the validity of the reconstruction, three different physical processes are numerically simulated in order to describe typical distributions of $\Psi_\phi$ that we can encounter in a flow of nanoparticles:\\
\begin{enumerate}
\item \emph{Uni-axial extensional flow}. The system of isotropically distributed particles at $z=0$ is stretched with a uni-axial extensional flow without any influence of rotary diffusion ($Pe=\infty$) until almost fully aligned at some downstream position $z$. On the centerline, there is thus a constant acceleration and the velocity is given by $w_c=\dot{\varepsilon}z$, with $\dot{\varepsilon}$ as an arbitrary constant extension rate and the alignment increases with $z$.
\item \emph{Rotary diffusion}. An almost fully aligned system of particles at $z=0$ is subject to rotary diffusion without any hydrodynamic forcing $(Pe=0)$. On the centerline, there is thus no acceleration and the velocity is given by $w_c=w_0$ with $w_0$ as an arbitrary constant velocity. The alignment will then decrease with $z$.
\item \emph{Equilibrium}. Given various constant $Pe\in]0,\infty[$, a system of isotropically distributed particles at $z=0$ is subject to a flow with constant acceleration ($w_c=\dot{\varepsilon}z$). Far downstream, as $z\rightarrow\infty$, the system will reach the equilibrium distribution given by the steady state solution $\Psi_\phi^\text{eq}$ given in eq.~(\ref{eq:equlibriumDistributions}). 
\end{enumerate}
In the cases 1~and~2, the ODF at a given downstream position $\Psi_\phi(z)$ is provided by numerically integrating the Smoluchowski equation (eq.~(\ref{eq:CylindricalSmoluchowski})) in Matlab R2013b. For case 3, the equilibrium ODF $\Psi_{\phi}^\text{eq}$ is given directly by eq.~(\ref{eq:equlibriumDistributions}) for various $Pe$. The azimuthal angle $\theta$ is distributed uniformly, i.e.~$\Psi_\theta=1$.

\begin{figure}
\includegraphics[width=0.6\textwidth]{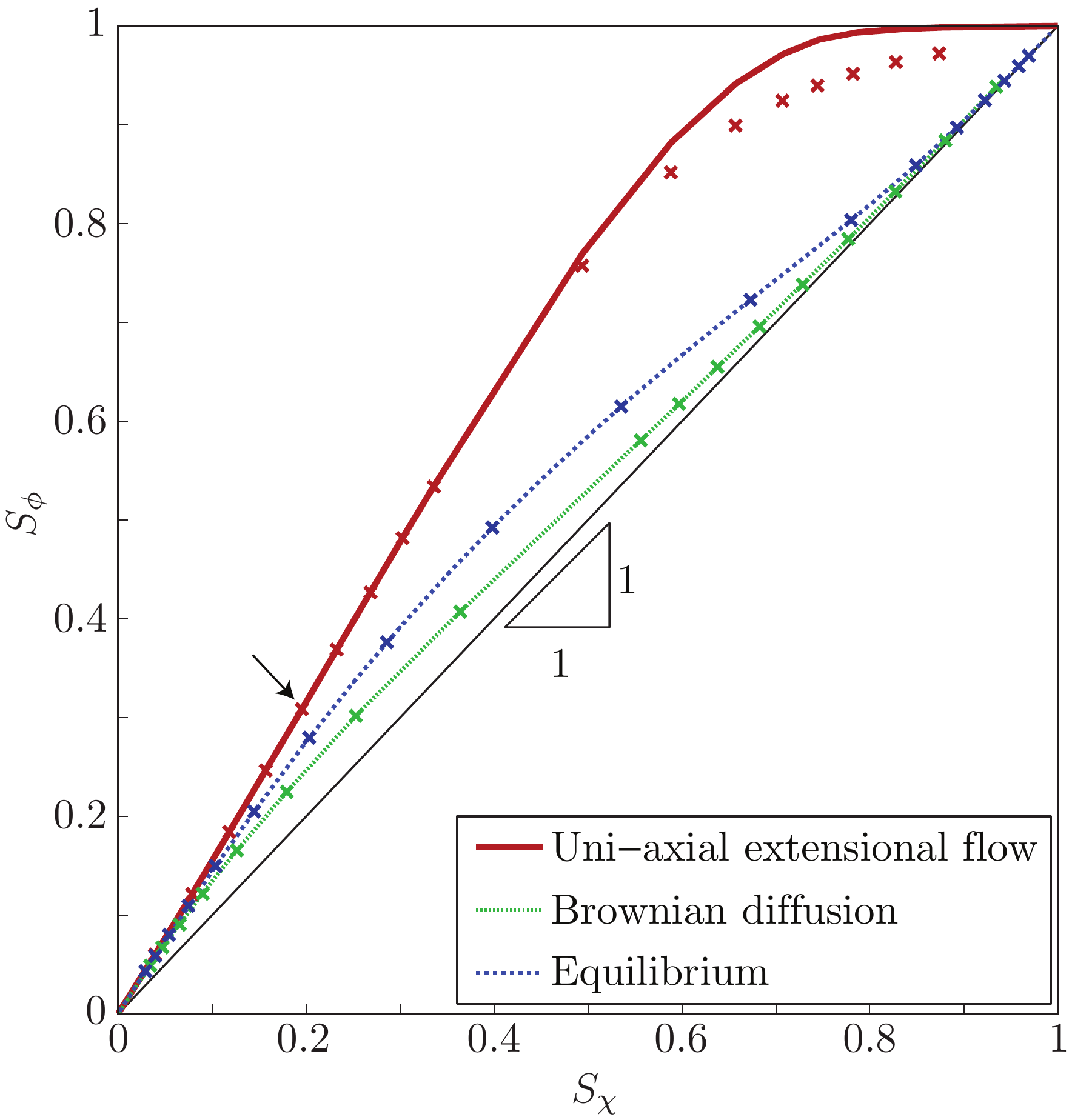}% Here is how to import EPS art
\caption{Illustration of how the measured projected order parameter $S_\chi$ differs from the true order parameter $S_\phi$ given the different flowing processes described in the text; the cross symbols denote the results from the reconstruction using $N_\text{LP}=10$; the arrow indicates the conditions for which the histograms in fig.~\ref{fig:Paper8angleHistograms} were drawn.}
  \label{fig:Paper8ValidationReconstruction}
\end{figure}

Given these ODFs, Monte-Carlo simulations were performed in Matlab R2013b. A total of $N=10^7$ particles are sampled, and for each particle, the orientation $(\phi,\theta)$ is randomly chosen from the distributions $\Psi_\phi$ and $\Psi_\theta$. For each particle, the projected angle $\chi$ is subsequently calculated through eq.~(\ref{eq:projectedAngle}), which is used to obtain the distribution $\Psi_\chi$ numerically. An example is shown in figure~\ref{fig:Paper8angleHistograms}. The distribution $\Psi_\phi$ in fig.~\ref{fig:Paper8angleHistograms}a is obtained from the simulation of the uni-axially extended system without diffusion at $S_\phi=0.3$. In order to account for the correct probability on the unit sphere, the distribution is multiplied with $|\sin\phi|$ and random numbers are generated from the distribution $\Psi_\phi|\sin\phi|$ to form the histogram in figure~\ref{fig:Paper8angleHistograms}a. Figure~\ref{fig:Paper8angleHistograms}b shows the histogram of the uniformly distributed values of $\theta$ and figure~\ref{fig:Paper8angleHistograms}c shows the corresponding histogram of the projected angle $\chi$ for all the particles, where the distribution is normalized according to eq.~(\ref{eq:3DNorm}) to form the ODF $\Psi_\chi^\text{(3D)}$.

The distribution $\Psi_\chi^\text{(3D)}$ is then used as an input for the reconstruction method described in the previous section. As seen in figure~\ref{fig:Paper8angleHistograms}a, the proposed method manages to almost perfectly recover the initial distribution of $\Psi_\phi$ when utilizing at least $N_\text{LP}=6$ Legendre polynomials.

In figure~\ref{fig:Paper8ValidationReconstruction}, it is observed how the order parameter $S_\phi$ will differ from the measured projected order parameter $S_\chi$ used for example by \citet{Hakansson_SAXSALIGN}. The discrepancy is larger during the acceleration of the flow when uni-axial extension is dominating, while approaching almost a one-to-one correspondence when Brownian rotary diffusion is dominating. Regardless of which type of distributions we encounter, we find that the proposed reconstruction manages to almost exactly recover $\Psi_\phi$ from $\Psi_\chi$ with $N_\text{LP}=10$. For the highly aligned ($S_\phi\gtrsim0.9$) uni-axially strained system, the reconstruction method is more sensitive and a larger number of Legendre polynomials would be needed. Note here also that $S_\phi\approx 0.9$ when $S_\chi\approx 0.6$. It would thus be very difficult to measure any higher alignment $S_\chi>0.6$ in a SAXS experiment with a uni-axially extended dispersion. A measured value of $S_\chi\approx 0.6$ is thus actually representing a system that is close to perfect alignment in reality.

\section{\label{sec:Paper8ApplSAXS} Application to real SAXS data}
The purpose of this study is to highlight and solve the problems that arises when comparing SAXS experiments with numerical simulations. Typically, the experiments provide the ODF $\Psi_\chi$ while the simulations provide $\Psi_\phi$. Depending of the physical process that drives an orientation distribution from isotropy to perfect alignment, or vice versa, there will be a different evolution of $\Psi_\phi$. A consequence of this is that we can measure the same projected order parameter $S_\chi$ at two points in the channel, but there might still be a difference in the alignment given by the true order parameter based on the polar angle $S_\phi$.

\begin{figure}
\includegraphics[width=0.99\textwidth]{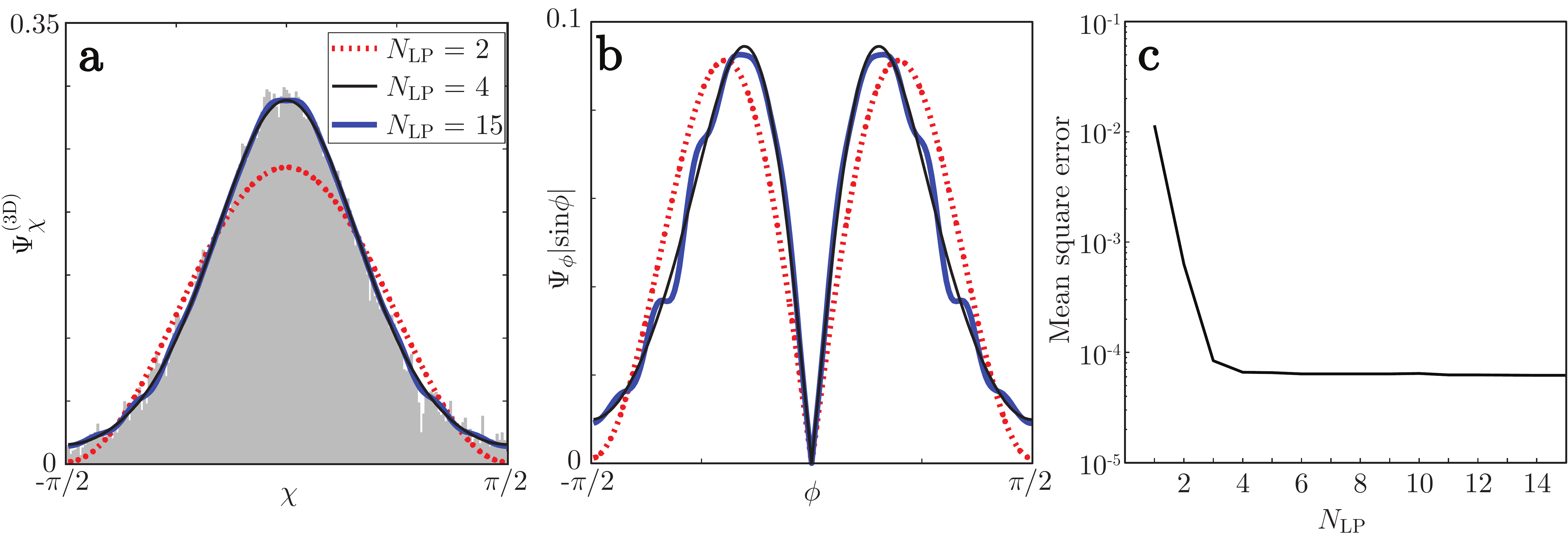}% Here is how to import EPS art
\caption{(a) The ODF $\Psi_\chi^\text{(3D)}$ obtained from SAXS experiments by \citet{Hakansson_NatComm} (gray histogram) is fitted with Legendre polynomials using $N_\text{LP}=2$, $4$ and $15$; (b) the resulting reconstruction of $\Psi_\phi$; (c) the mean square error between the experiment and the model of $\Psi_\chi^\text{(3D)}$ as function of $N_\text{LP}$.}
  \label{fig:ChoosingNLP}
\end{figure}

To demonstrate the application of the present reconstruction, the original SAXS distributions $\Psi_\chi$ from \citet{Hakansson_NatComm} are used. The choice of $N_\text{LP}$ must be done with care due to scatter in the experimental data. Figure~\ref{fig:ChoosingNLP}a shows $\Psi_\chi^\text{(3D)}$ obtained at $z=2.5h$ ($h$ is the side of the quadratic cross-section of the flow-focusing device) from the SAXS experiments by \citet{Hakansson_NatComm}. Choosing $N_\text{LP}=2$ (red dashed curve) to fit $\Psi_\chi^\text{(3D)}$ is not sufficient, but on the other hand choosing $N_\text{LP}=15$ (blue thick solid curve) creates some "wiggles" in the fitted curve as the scatter in the data influences the fitting. This is even more apparent when looking at the resulting distribution of $\Psi_\phi$ in figure~\ref{fig:ChoosingNLP}b. When plotting the mean square error between the experimental data of $\Psi_\chi^\text{(3D)}$ and the fitted curve as function of $N_\text{LP}$ in figure~\ref{fig:ChoosingNLP}c it is clear that there really is no improvement of the fitting above $N_\text{LP}>4$. We therefore implement a strategy to avoid overfitting by setting a stopping criteria based on the mean square error. The details are given in the supplementary information. 

\begin{figure}
\includegraphics[width=0.99\textwidth]{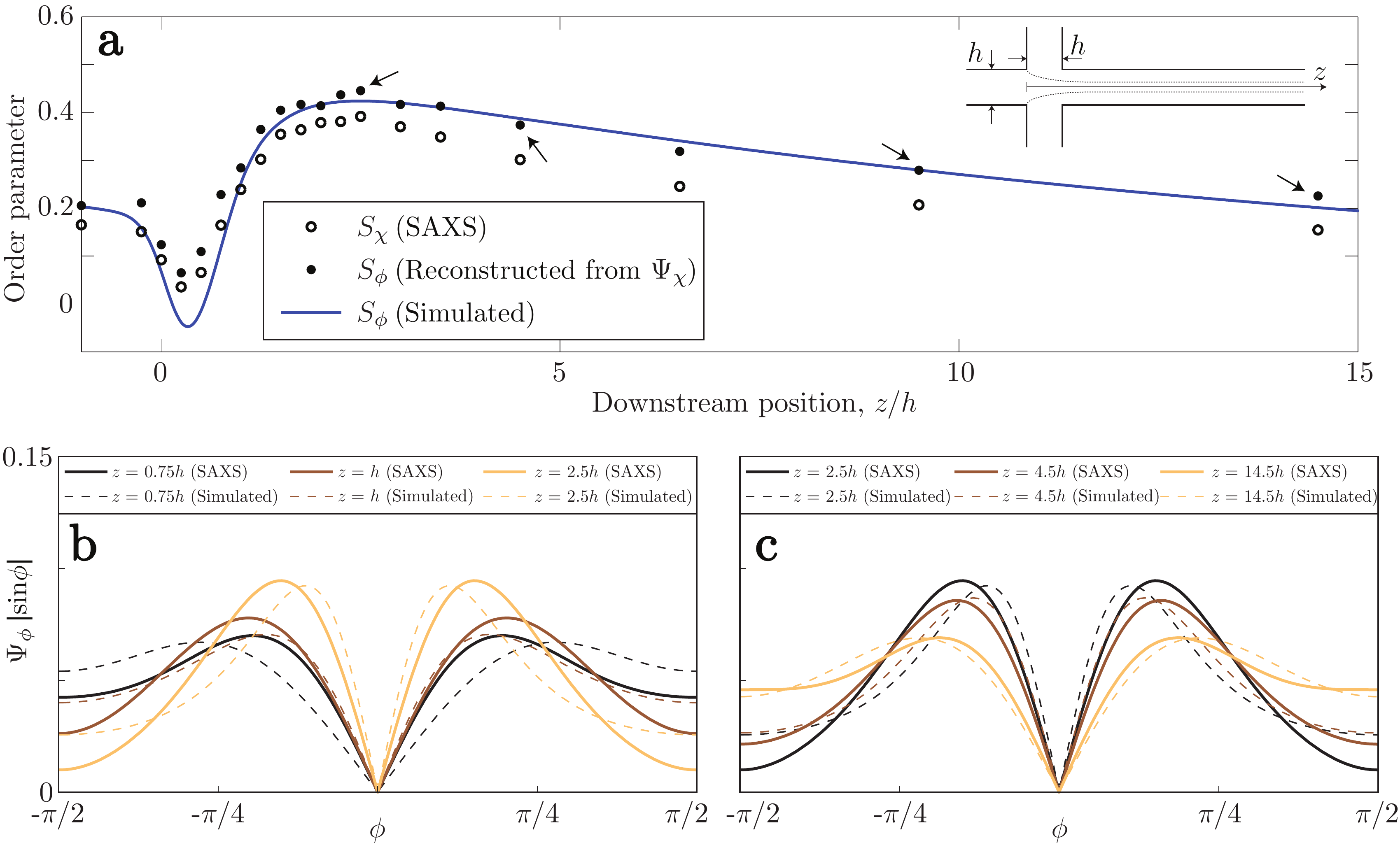}% Here is how to import EPS art
\caption{Results of the reconstructed order parameter $S_\phi$ from the SAXS measurements by \citet{Hakansson_NatComm}; the results are compared with the simulations using the Smoluchowski equation (eq.~(\ref{eq:CylindricalSmoluchowski})); (a) order parameter as function of downstream position $z$; the arrows indicate the positions that are used for the numerical drying experiments in figure~\ref{fig:Paper8numDryingResults}; (b), (c) the evolution of the ODFs $\Psi_\phi$ at positions in the acceleration region of the channel and at positions after the focusing region.}
  \label{fig:Paper8_SAXS_application}
\end{figure}

Using this strategy to find $N_\text{LP}$, the reconstruction method is applied to the original data by \citet{Hakansson_NatComm}. Figure~\ref{fig:Paper8_SAXS_application}a shows how the reconstructed $S_\phi$ differs from the value of $S_\chi$. The value of $S_\phi$ is consistently higher than $S_\chi$ with a difference of around 0.05. Even though the alignment is mainly through uni-axial extension, the difference between $S_\phi$ and $S_\chi$ is much smaller than the prediction from the uni-axial extension curve in figure~\ref{fig:Paper8ValidationReconstruction}. This is most probably attributed to substantial rotary diffusion during the extension process. Consequently, the order parameter by \citet{Hakansson_NatComm} is actually a decent estimate of $S_\phi$.

Now, a comparison can be done with simulations using the Smoluchowski equation in eq.~(\ref{eq:CylindricalSmoluchowski}). As a boundary condition to the simulations, the ODF $\Psi_\phi$ at $z=-h$ is set to be the reconstructed ODF from \citet{Hakansson_NatComm} at the same position. The centerline velocity $w_c(z)$ for the same flow conditions is taken from \citet{Hakansson_SAXSALIGN}, which was obtained numerically through the two-fluid level set method with a core flow with kinematic viscosity of $\eta=40$~mPa~s and sheath flow of water. This centerline velocity was also verified experimentally through micro particle tracking velocimetry ($\mu$PTV) by \citet{Hakansson_SAXSALIGN}. Further parameters used in the present simulations are $\Lambda\approx 1$ and $D_r=0.23$~rad$^2$/s, where the latter is chosen to match the experimental data as well as possible.

The simulated results are illustrated with the blue line in figure~\ref{fig:Paper8_SAXS_application}, where the order parameter $S_\phi$ is plotted versus the downstream position $z$. Although the model is not perfectly capturing the experiments, the qualitative appearance is fairly good and it could be potentially seen as a good model for the physical system. However, comparing the actual evolution of the ODFs at different downstream positions in figures ~\ref{fig:Paper8_SAXS_application}b-c, it is found that the model does not perform very well in capturing the shape of the ODFs. This is of course not very surprising, since the model chosen for this particular comparison in eq.~(\ref{eq:CylindricalSmoluchowski}) relies on many questionable assumptions; one of the more crucial being that the CNF dispersion is dilute and that fibrils do not interact. For future work, in order to improve the Smoluchowski model to account for concentration effects, additional modifications are needed, for example introducing a rotary diffusion coefficient that is dependent both on the concentration and the local ODF as done by \citet{Hakansson_SAXSALIGN}.

\section{\label{sec:Paper8ApplDrying} Application to the drying process}
In the process described by \citet{Hakansson_NatComm}, the aligned dispersion of CNF is locked in an arrested (gel) state when exiting the flow-focusing device. The gel thread is subsequently dried afterwards to form the final filament. During the removal of water from the gel, the cylindrical thread goes from a radius of approximately 0.5~mm to 20~$\mu$m while maintaining the same length. The thread has thus shrunken radially with a ratio of $\beta=20/500=0.04$. During the drying of the CNF gel, the radial component of the fibril orientation (in the cylindrical geometry given by the gel thread) is assumed to also decrease with the same shrinking ratio $\beta$. The result is that the $z$-component of the fibril orientation increases, and thus also the fibril alignment. This radial shrinking process will be used to simulate how the fibril alignment increases during the drying of the gel thread. The detailed mathematical description of the process is given in the supplementary information and a movie illustrating how fibrils align during this process is provided in the online supplementary material.

In the numerical drying simulations, it will be assumed that the gel-transition will lock the ODF in the dispersion at a certain downstream position $z$. Therefore, the reconstructed ODF $\Psi_{\phi}$ from the SAXS experiments at four different positions after the focusing section (pointed out with the small arrows in fig.~\ref{fig:Paper8_SAXS_application}a) are used as initial ODF for the numerical drying simulations. These positions are $z=2.5h$,~$4.5h$,~$9.5h$ and $14.5h$. The initial orientations $\phi_0$ of the particles are randomly chosen from these ODF exactly as done previously. The initial angle $\theta_0$ is again chosen from a uniform distribution. The system is then aligned by decreasing $\beta$ from 1 to 0 to simulate the drying process.

\begin{figure}[t]
\includegraphics[width=0.99\textwidth]{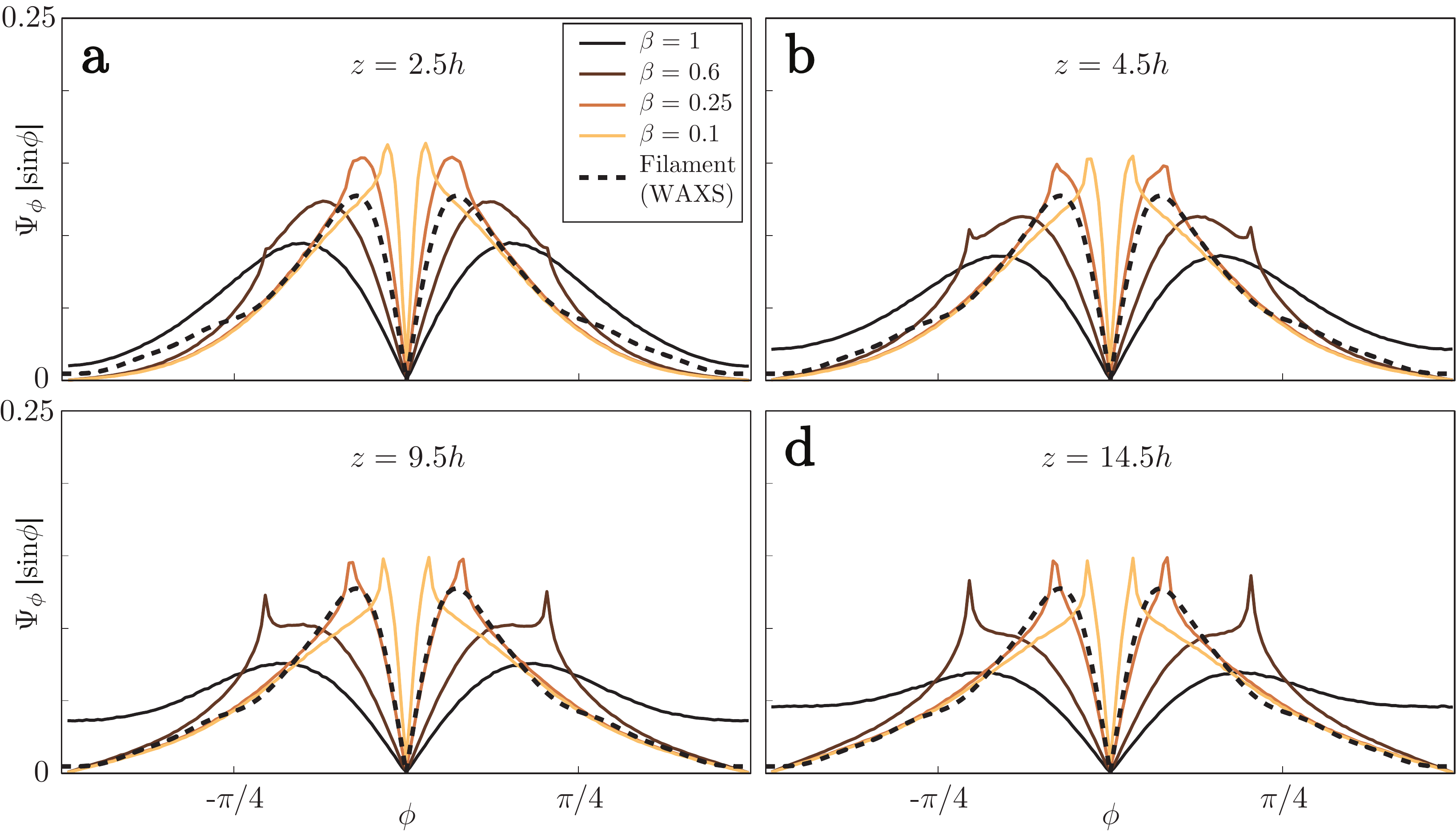}% Here is how to import EPS art
\caption{Results from the simulated drying process using initial ODFs $\Psi_\phi$ from four different positions: (a)~$z=2.5h$, (b)~$4.5h$, (c)~$9.5h$ and (d)~$14.5h$ obtained by reconstructing the ODFs from the SAXS experiments by \citet{Hakansson_NatComm}; the results are compared with the actual WAXS data of the dried cellulose filament in \citet{Hakansson_NatComm}.}
  \label{fig:Paper8numDryingResults}
\end{figure}

The resulting ODFs after the numerical drying simulations are illustrated in figure~\ref{fig:Paper8numDryingResults} for different shrinking ratios $\beta$ at the given downstream positions. As expected, assuming the drying process to be represented well by the radial shrinking principle, the alignment can increase substantially during the process. For example the initial distribution at $z=14.5h$ goes from an order parameter of $S_\phi\approx 0.2$ to $S_\phi\approx 0.6$ during drying. As a matter of fact, even a completely random distribution ($S_\phi=0$) is found to reach an alignment of $S_\phi=0.5$ during this process (shown as a supplementary movie). As the shrinking process starts, the fibrils oriented in the $\phi=\pm\pi/2$-plane are aligning quicker than the fibrils with lower $|\phi|$. The result is that there will be a sharp peak visible in the ODF. Assuming that there is some rotary diffusion present also during the drying process, either due to Brownian motion or fibril contacts, this peak will probably be smoothed out in reality.

To assess the validity of this principle, a comparison is made with the dried cellulose filament produced through the given flow conditions, which was presented by \citet{Hakansson_NatComm}. An ODF of the fibrils inside the filament was obtained through WAXS, which means that the presented ODF is a measurement of $\Psi_\chi$. The alignment was found to be $S_\chi=0.5$ and representing a projected mean fibril angle of $\langle \chi \rangle = \arccos \sqrt{(2S_\chi+1)/3}=35^\circ$. Assuming that the ODF of the azimuthal angle $\theta$ is uniform also in the filament, the ODF $\Psi_\phi$ is reconstructed in the same way as for the SAXS data. The order parameter with respect to the polar angle $\phi$ is found to be $S_\phi=0.61$, with a corresponding mean fibril angle of $\langle \phi \rangle = \arccos \sqrt{(2S_\phi+1)/3}=31^\circ$. The alignment is thus actually higher than what was presented by \citet{Hakansson_NatComm}. However, even though the value of $S_\phi$ was substantially ($\approx 20\%$) higher, there was only a small decrease ($\approx 11\%$) of the mean fibril angle.

The data of the dried filament by \citet{Hakansson_NatComm} is illustrated with the dashed curves in figures~\ref{fig:Paper8numDryingResults}a-d. Observing first the simulated ODF from the position of maximum alignment ($z=2.5h$) in figure~\ref{fig:Paper8numDryingResults}a, it is seen that the ODF does not approach the distribution of the filament with decreasing $\beta$. 
Interestingly, the ODF of the dried cellulose filament obtained through WAXS seems to be very close to the numerical drying simulations at $z=9.5h$ and $z=14.5h$ for $\beta=0.25$, assuming that the peak in the ODF is smoothed out. The close matching of the ODFs at least strengthens the hypothesis that the drying process is mainly governed by the radial shrinking principle. The indication is also that the gelling in this case occurs somewhere around $z\approx 10h$. This is also in good agreement with the results by \citet{Hakansson_RSC}. The results also suggest that we could theoretically produce a filament with an order parameter of $S_\phi\approx 0.7$ and $\langle \phi \rangle = 27^\circ$, by just controlling the gelling at the present flow conditions to occur at the position of maximum alignment, i.e.~at $z=2.5h$. Of course, there are practical limitations to achieving this value as it also takes some time for the gel to form.

In conclusion, it is found here that a shrinking ratio of $\beta\approx 0.25$ seems to be enough to almost obtain the true ODF of the dried filament. In reality the expected shrinking ratio is however close to $\beta=0.04$. The reason is believed to be that the true drying process probably is a lot more intricate than this naive model, especially in the final stages of the drying. As the particle concentration increases towards the end of the drying ($\beta < 0.25$), the particles might sterically hinder each other from aligning according to our simplified model. Due to the results in this work, it is therefore hypothesized that the ODF remains almost constant during the final stages of the drying at $\beta=0.25\rightarrow 0.04$. 

\section{\label{sec:Paper8concl} Summary and conclusions}
Measuring the alignment of elongated nanoparticles (nanofibrils) is often done using small angle X-ray scattering techniques (SAXS), where the beam direction is perpendicular to the flow direction. It is however known that the anisotropic scattering resulting from an aligned system of particles reflects the anisotropy of the projected angle $\chi$ of the fibrils in the viewing plane perpendicular to the beam direction. Therefore, the typical measurement can only provide the orientation distribution function (ODF) of the projected angle $\Psi_\chi$.

In this work, we present a new way of reconstructing the three-dimensional ODF $\Psi_\phi$ (where $\phi$ is the polar angle between the fibril and the flow direction) by using $\Psi_\chi$. The additional assumption is that the distribution of the azimuthal angle $\theta$ around the flow direction is uniform, i.e.~$\Psi_\theta$ is constant; an assumption that is approximately valid in dispersed particle flows through cylindrical geometries. The ODF $\Psi_\phi$ can then be used to compare with simulations of dispersed nanoparticles in order to understand more of the underlying physics. This method furthermore opens up the possibility to access the three-dimensional orientation using SAXS without having to rotate the sample, given that the assumption of cylindrical symmetry is valid.

The reconstruction method was applied to revise the experimental results by \citet{Hakansson_NatComm}, who studied the hydrodynamic alignment and assembly of cellulose nanofibrils (CNF) in a flow-focusing device to produce continuous cellulose filaments. It is found in the present work that the three-dimensional alignment (represented by the order parameter $S_\phi$) is slightly higher than the previously presented projected alignment (represented by $S_\chi$). A simple model to simulate the evolution of $\Psi_\phi$ in the flow-focusing device is applied by using the Smoluchowski equation, assuming no fibril interactions. Even though the evolution of $S_\phi$ versus downstream position could be qualitatively captured with this simple model, it is obvious that it does not simulate the true evolution of $\Psi_\phi$ as measured with SAXS.

Finally, the reconstruction was also applied to the ODF obtained through wide angle X-ray scattering (WAXS) experiments of a dried filament by \citet{Hakansson_NatComm}. It was found that the results were consistent with the assumption that the drying process is governed by a radial shrinking process as water is removed from the gel thread. To further verify this simplified model of the drying process, additional in-situ SAXS/WAXS experiments can be conducted in the future, where the structure is studied during the actual drying process similar to the study by \citet{sen2007slow}.

The results in this work opens up new possibilities to study elongated nanoparticles in cylindrical flows and quantitatively compare experiments with numerical simulations. This will potentially lead to improved models and reveal the true mechanisms behind the angular dynamics of the dispersed nanoparticles. The future improved models should then be assessed by comparing the actual three-dimensional ODF in the channel rather than comparing integrated order parameters.

Future research is also suggested in the direction of assessing the validity of a constant $\Psi_\theta$. This can possibly be found by coupling the full three-dimensional version of the Smoluchowski equation to two-fluid simulations in the three-dimensional geometry. If $\Psi_\theta$ can not be assumed to be constant, the reconstruction method presented in this work will fail and it would require other considerations to obtain $\Psi_\phi$ from a measurement of $\Psi_\chi$. However, recent advances in 3D tomographic SAXS\cite{schaff2015six,skjonsfjell2016} could possibly be used also for flowing dispersions of nanoparticles. Using this technique, both $\Psi_\phi$ and $\Psi_\theta$ could possibly be obtained at a given position in space.

\begin{acknowledgement}
The authors would like to thank Dr. Karl H\aa kansson for sharing the data needed to demonstrate the present reconstruction method. Dr. Anders Dahlkild is also greatly thanked for his insightful comments regarding orientation distribution functions. The authors furthermore acknowledge financial support from the Wallenberg Wood Science Center (WWSC).
\end{acknowledgement}

\section{Supplementary Material}
The following files can be downloaded from \url{https://www.mech.kth.se/~rosen/MCSAXS/}:
\begin{enumerate}
\item Matlab code for reconstructing $\Psi_\phi$ from an input ODF $\Psi_\chi$ using $N_\text{LP}\in[1,15]$.
\item Movie showing how isotropically distributed particles align during a radial shrinking process.
\end{enumerate}

\bibliography{REFERENCES}% Produces the bibliography via BibTeX.

\pagebreak

\begin{center}
\section{Supplementary information: Evaluating alignment of elongated nanoparticles in cylindrical geometries through small angle X-ray scattering experiments}
\end{center}
~\\

\section{Equilibrium solution to the Smoluchowski equation}
Assuming the orientation distribution function (ODF) $\Psi$ to only be dependent on the polar angle $\phi$, the stationary Smoluchowski equation is given by:\\
\begin{equation}
w_c\frac{\partial \Psi_\phi}{\partial z}=\frac{1}{\sin{\phi}}\frac{\partial}{\partial \phi}\left(D_r\sin\phi\frac{\partial \Psi_\phi}{\partial \phi}+\frac{3}{2}\dot{\varepsilon}\Lambda\cos\phi\sin^2\phi\Psi_\phi\right).
\end{equation}
~\\
Given a constant $D_r$, $\Lambda$ and $\dot{\varepsilon}$, the solution approaches an equilibrium ODF $\Psi_\phi^\text{eq}$ at $z\rightarrow \infty$. This is obtained by the solution of $\partial \Psi_\phi/\partial z=0$, i.e.~when\\
\begin{equation}
\frac{\partial}{\partial \phi}\left(D_r\sin\phi\frac{\partial \Psi_\phi^\text{eq}}{\partial \phi}+\frac{3}{2}\dot{\varepsilon}\Lambda\cos\phi\sin^2\phi\Psi_\phi^\text{eq}\right)=0.
\end{equation}
~\\
Integrating both sides with respect to $\phi$, we obtain\\
\begin{equation}
\sin\phi\left(D_r\frac{\partial \Psi_\phi^\text{eq}}{\partial \phi}+\frac{3}{2}\dot{\varepsilon}\Lambda\cos\phi\sin\phi\Psi_\phi^\text{eq}\right)=C, 
\end{equation}
~\\
with the integration constant $C$. In order for this to be valid for all values of $\phi$, the constant $C$ must equal to zero, which leads to the following ordinary differential equation of the first order:\\
\begin{equation}
\frac{\partial \Psi_\phi^\text{eq}}{\partial \phi}+\frac{3}{2}\underbrace{\frac{\dot{\varepsilon}\Lambda}{D_r}}_{=Pe}\cos\phi\sin\phi\Psi_\phi^\text{eq}=0.
\end{equation}
~\\
The solution of this ODE is\\
\begin{equation}
\Psi_{\phi}^\text{eq}=A\exp(-\frac{3}{4}Pe\sin^2 \phi),
\end{equation}
where the constant $A$ is found through the normalization
\begin{equation}
\int_0^{2\pi}d\theta\int_{-\pi/2}^{\pi/2} \Psi_{\phi}^\text{eq}|\sin \phi| \textrm{d}\phi=1.
\end{equation}

\section{Theory behind the reconstruction}

\begin{figure}
\includegraphics[width=0.3\textwidth]{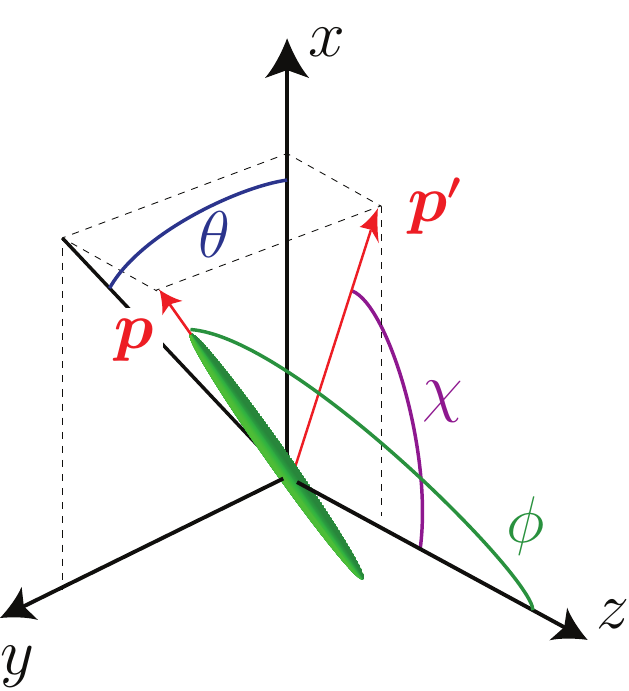}% Here is how to import EPS art
\caption{Same figure as fig.~1a in the main manuscript; definition of the orientation of an elongated particle with unit vector $\boldsymbol{p}$ along the major axis; the polar angle of the particle to the $z$-direction is denoted with $\phi$ and the corresponding azimuthal angle in the $xy$-plane is $\theta$; the projection of the unit vector $\boldsymbol{p}$ on the viewing ($xz$) plane is denoted $\boldsymbol{p}'$ and has the angle $\chi$ with respect to the $z$-axis.}
  \label{fig:ParticleDef}
\end{figure}

Knowing the Legendre decomposition of $\Psi_\phi$, it is possible to project the distribution in the $xz$-plane, to obtain a bi-dimensional distribution $\Psi^\text{(2D)}_\chi(\chi)$. The unit vector $\boldsymbol{p}=(p_x,p_y,p_z)$ giving the particle orientation is projected in the $xz$-plane into the non-unit vector\\
\begin{equation}
\boldsymbol{p'}=(p_x,p_z)=(\sin \phi \cos \theta, \cos \phi)=|\boldsymbol{p'}|(\sin \chi, \cos \chi),
\end{equation}
~\\
where $|\boldsymbol{p'}|\equiv p' =\sqrt{\sin^2 \phi \cos^2 \theta + \cos^2 \phi}$. That the two expressions of $\boldsymbol{p'}$ must be the same is seen easily in figure~\ref{fig:ParticleDef}. The function $\Psi_\phi$ can be decomposed into the Legendre polynomials $P_j(\cos\phi)$ according to\\
\begin{equation}
\Psi_\phi=\sum_{j=0,2,4,...}^{2(N_\text{LP}-1)}\frac{2j+1}{4\pi} \langle P_j \rangle_\phi P_j(\cos \phi),
\label{eq:legDecomp}
\end{equation}
~\\
which means that $\Psi_\phi$ only depends on $\cos \phi = p_z$. The projection is made by integrating the ODF in the $y$-direction\\
\begin{equation}
\Psi^\text{(2D)}_{\boldsymbol{p'}}=\int_{-\infty}^{+\infty} \textrm{d}p_y \Psi_\phi(p_z) \delta(1-p),
\label{eq:Psi2D_w_deltaFunct}
\end{equation}
~\\
where $p=\sqrt{p_x^2+p_y^2+p_z^2}=\sqrt{p_y^2+p'^2}$ and the Dirac function ensures that the integration is only done for vectors $\boldsymbol{p}$ that fulfill the requirement $|\boldsymbol{p}|\equiv p=1$ since $\boldsymbol{p}$ is a unit vector. This Dirac function can be rewritten using the relationship\\
\begin{equation}
\delta(f(p_y))=\frac{\sum_{p_{y,0}}\delta(p_y-p_{y,0})}{|f'(p_{y,0})|}
\label{eq:deltaFunctRelation}
\end{equation}
~\\
where $p_{y,0}$ is a real root to the function $f(p_y)$. This is applied to our case where \\
\begin{eqnarray}
f(p_y)&=&1-\sqrt{p_y^2+p'^2},\\
f'(p_y)&=&-\frac{p_y}{\sqrt{p_y^2+p'^2}}.
\end{eqnarray}
~\\
The function $f(p_y)$ has two roots at $p_{y,0}=\pm\sqrt{1-p'^2}$, which means that the delta function in eq.~(\ref{eq:Psi2D_w_deltaFunct}) can be rewritten using the relation~(\ref{eq:deltaFunctRelation}) as\\
\begin{equation}
\delta(1-p) = \frac{\delta(p_y-\sqrt{1-p'^2})+\delta(p_y+\sqrt{1-p'^2})}{\sqrt{1-p'^2}}.
\end{equation}
~\\
The integral in eq.~(\ref{eq:Psi2D_w_deltaFunct}) thus becomes\\
\begin{equation}
\Psi^\text{(2D)}_{\boldsymbol{p'}}=\int_{-\infty}^{+\infty} \textrm{d}p_y \Psi_\phi(p_z) \frac{\delta(p_y-\sqrt{1-p'^2})+\delta(p_y+\sqrt{1-p'^2})}{\sqrt{1-p'^2}},
\label{eq:Psi2D_w_deltaFunct2}
\end{equation}
which has the solution\\
\begin{equation}
\Psi^\text{(2D)}_{\boldsymbol{p'}}=\frac{2 \Psi_\phi(p_z) }{\sqrt{1-p'^2}},
\end{equation}
~\\
where $\Psi^\text{(2D)}_{\boldsymbol{p'}}$ depends on $p'\in[0,1]$ and $\chi\in[-\pi,\pi]$. By symmetry, the $\chi$ range can be reduced to $[-\pi/2, \pi/2]$ and $\Psi^\text{(2D)}_{\boldsymbol{p'}}$ multiplied by a factor $2$. To obtain the 2D ODF $\Psi^\text{(2D)}_\chi$ as a function of $\chi$ only, an integration is made over $p'$ since $p_z=\cos\phi=p'\cos\chi$:\\
\begin{equation}
\Psi^\text{(2D)}_\chi=\int_{0}^{1} \frac{4 p' \textrm{d}p' \Psi_\phi(p' \cos \chi)}{\sqrt{1-p'^2}}.
\label{eq:Psi2DIntegral}
\end{equation}
~\\
The ODF $\Psi_\phi$ can be decomposed into Legendre polynomials $P_j(\cos\phi)$ according to eq.~(\ref{eq:legDecomp})\\
\begin{equation}
\Psi_\phi=\sum_{j=0,2,4,...}^{2(N_\text{LP}-1)}\frac{2j+1}{4\pi} \langle P_j \rangle_\phi P_j(\underbrace{p'\cos\chi}_{\cos\phi}),
\end{equation}
~\\
Using this, the integral in eq.~(\ref{eq:Psi2DIntegral}) can be re-written as\\
\begin{equation}
\Psi^\text{(2D)}_\chi=\sum_{j=0,2,4,...}^{2(N_\text{LP}-1)}\frac{2j+1}{4\pi} \langle P_j \rangle_\phi \int_{0}^{1} \frac{4 p' \textrm{d}p' P_j(p' \cos \chi)}{\sqrt{1-p'^2}}.\label{eq:legendre_decomposition_chi_2D}
\end{equation}
~\\
This ODF is normalized according to\\
\begin{equation}
\int_{-\pi/2}^{\pi/2} \Psi^\text{(2D)}_\chi \textrm{d}\chi =1.
\end{equation}

We can now consider this 2D ODF as a 3D ODF by changing only the normalization according to\\
\begin{equation}
\Psi^\text{(3D)}_\chi =\underbrace{\left(\int_0^{2\pi}d\theta\int_{-\pi/2}^{\pi/2} \Psi^\text{(2D)}_\chi |\sin \chi| \textrm{d}\chi \right)^{-1}}_{\alpha}\Psi^\text{(2D)}_\chi\equiv \alpha \Psi^\text{(2D)}_\chi.
\end{equation}
~\\
Now, it is possible to decompose $\Psi^\text{(3D)}_\chi$ into Legendre polynomials:\\
\begin{equation}
\Psi^\text{(3D)}_\chi=\sum_{i=0,2,4,...}^{2(N_\text{LP}-1)}\frac{2i+1}{4\pi} \langle P_i \rangle_\chi P_i(\cos \chi),
\label{eq:legendre_decomposition_chi}
\end{equation}
~\\
where the order parameters are given by\\
\begin{eqnarray}
\langle P_i \rangle_\chi &= & \int_0^{2\pi}d\theta\int_{-\pi/2}^{\pi/2} \Psi^\text{(3D)}_\chi P_i(\cos \chi) |\sin \chi| \textrm{d}\chi\\
&=& 2\pi \alpha\sum_{j=0,2,4,...}^{2(N_\text{LP}-1)}\frac{2j+1}{4\pi}   \langle P_j \rangle_\phi  C_{i,j},\label{eq:matrix_C}
\end{eqnarray}
~\\
using\\
\begin{equation}
C_{i,j}=\int_{0}^{1} \frac{4 p' \textrm{d}p' }{\sqrt{1-p'^2}} \int_{-\pi/2}^{\pi/2}  P_i(\cos \chi) P_j(p' \cos \chi) |\sin \chi| \textrm{d}\chi.
\end{equation}
~\\
We now have a relationship between the order parameters $\langle P_i \rangle_\chi$ and $\langle P_j \rangle_\phi$ via an $N_\text{LP}\times N_\text{LP}$ matrix.

\section{Choosing $N_\text{LP}$}
In order to choose a fit experimental data with as few Legendre polynomials as possible while still having a low error, i.e.~to not have overfitting of the data, we apply the following strategy:\\
\begin{enumerate}
\item Start with $N_\text{LP}=2$.
\item Calculate the mean square error $\varepsilon(N_\text{LP})$ between the experimental data of $\Psi^\text{(3D)}_\chi$ and the corresponding fit.
\item Calculate the mean square error $\varepsilon(N_\text{LP}-1)$ between the experimental data of $\Psi^\text{(3D)}_\chi$ and the corresponding fit.
\item If $\log_{10}\varepsilon(N_\text{LP}-1)-\log_{10}\varepsilon(N_\text{LP})>0.05$, start over at step 1 with $N_\text{LP}+1$.
\item If $\log_{10}\varepsilon(N_\text{LP}-1)-\log_{10}\varepsilon(N_\text{LP})<0.05$, choose the value $N_\text{LP}-1$ for the reconstruction.
\end{enumerate}

\section{Radial shrinking model}

\begin{figure}
\includegraphics[width=0.99\textwidth]{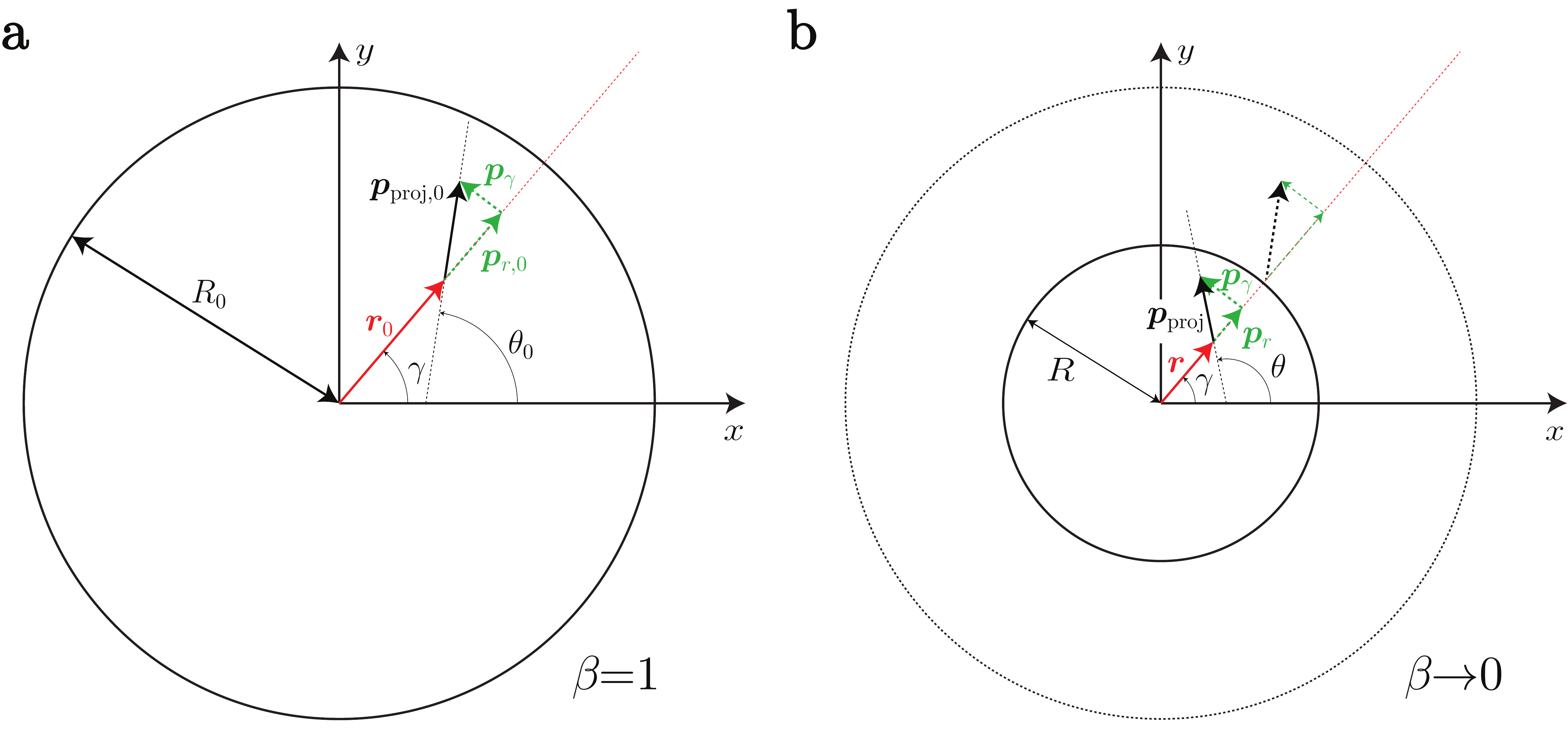}% Here is how to import EPS art
\caption{Illustration the numerical shrinking experiment; (a) initially we consider a circular cross-section of the cylindrical geometry with radius $R_0$; a particle is place at polar coordinates ($|\boldsymbol{r}_0|,\gamma$) and orientation $(\phi_0,\theta_0)$ with the projected unit symmetry axis denoted $\boldsymbol{p}_{\text{proj,0}}$; (b) after the shrinking with factor $\beta$, the circular cross-section is reduced to a radius $\beta R_0$, the position of the particle is moved to $\beta\boldsymbol{r}_0$ and the radial component of the orientation is reduced to $\beta p_{r,0}$.}
  \label{fig:Paper8numShrinkingPrinciple}
\end{figure}

Consider a circular cross-section of a cylindrical flow of radius $R$. A particle is positioned at a certain position vector $\boldsymbol{r}_0$, which in polar coordinates is described with the distance from the center $|\boldsymbol{r}_0|$ and the angle $\gamma$ towards the $x$-direction as shown in figure~\ref{fig:Paper8numShrinkingPrinciple}a. The unit vectors of the cylindrical coordinates are given by $\boldsymbol{e}_r=(\cos\gamma,\sin\gamma)$ and $\boldsymbol{e}_\gamma=(-\sin\gamma,\cos\gamma)$. The particle has a certain orientation $\phi_0$ and $\theta_0$ and the projected unit direction of the particle $\boldsymbol{p}_{\text{proj},0}$ on the $xy$-plane is given by $\boldsymbol{p}_{\text{proj},0}=(p_{x,0},p_{y,0})=(\sin\phi_0\cos\theta_0,\sin\phi_0\sin\theta_0)$. The radial and azimuthal components of this projected vector are given by $p_{r,0}=\boldsymbol{p}_{\text{proj},0}\cdot\boldsymbol{e}_r$ and $p_\gamma=\boldsymbol{p}_{\text{proj},0}\cdot\boldsymbol{e}_\gamma$, respectively. The general relationships between these components and the particle orientation is given by\\
\begin{equation}
\sin\phi_0=\sqrt{p_{r,0}^2+p_\gamma^2}, \qquad \cos\theta_0=\frac{\boldsymbol{p}_{\text{proj},0}\cdot\boldsymbol{e}_x}{\sqrt{p_{r,0}^2+p_\gamma^2}}.
\label{eq:Paper8q}
\end{equation}
~\\
To simulate how the alignment changes as the sample is drying, we will assume that everything is shrinking radially. Going from a circular cross-section of radius $R_0$ to one with radius $R$, we will also change\\
\begin{equation}
\frac{R}{R_0}=\frac{|\boldsymbol{r}|}{|\boldsymbol{r}_0|}=\frac{p_r}{p_{r,0}}=\beta,
\label{eq:Paper8shrinking}
\end{equation}
~\\
where $\beta$ is defined as the shrinking ratio. During the drying, we then assume $\gamma$ and $p_\gamma$ to remain unchanged. The result is that the projected length of the symmetry axis $\boldsymbol{p}_\text{proj}$ decreases, i.e.~the distribution of particles become more aligned along the $z$-direction. This process is illustrated in figure~\ref{fig:Paper8numShrinkingPrinciple}b. The particle positions are initially uniformly distributed on the circular cross section. Given a shrinking ratio $\beta$ and an initial distribution of $\theta_0$ and $\phi_0$, the resulting angles $\theta$ and $\phi$ after shrinking are calculated using the relations:\\
\begin{equation}
\phi=\arcsin\sqrt{(p_{r,0}\beta)^2+p_\gamma^2},
\label{eq:Paper8PhiAfterShrinking}
\end{equation}
\begin{equation}
\theta=\arccos\left(\frac{(\beta p_{r,0}\boldsymbol{e}_r+p_\gamma\boldsymbol{e}_\gamma)\cdot\boldsymbol{e}_x}{\sqrt{(\beta p_{r,0})^2+p_\gamma^2}}\right).
\label{eq:Paper8ThetaAfterShrinking}
\end{equation}

\end{document}